\documentclass[11pt,a4paper]{article} 
\pdfoutput=1
\usepackage{jcappub}

\title{Supercurvaton}
\author[a]{Vittoria Demozzi,} \author[b]{Andrei Linde} \author[a]{and Viatcheslav Mukhanov}

\affiliation[a]{Arnold-Sommerfeld-Center for Theoretical Physics, Department
f\"ur Physik, Ludwig-Maximilians-Universit\"at M\"unchen, Theresienstr. 37,
D-80333, Munich, Germany} \affiliation[b]{Stanford Institute for Theoretical Physics and Department of Physics, \\ Stanford University, Stanford, CA 94305, USA} 
\emailAdd{vittoria.demozzi@physik.uni-muenchen.de} \emailAdd{alinde@stanford.edu} \emailAdd{Viatcheslav.Mukhanov@physik.uni-muenchen.de}
\abstract{We discuss observational consequences of the curvaton scenario, which
naturally appears in the context of the simplest model of chaotic inflation
in supergravity. The non-gaussianity parameter $f_{\mathrm{NL}}$ in this
scenario  can take values in the observationally interesting range from $O(10)$ to $O(100)$. These values may be different in different parts of the universe. The regions where $f_{\mathrm{NL}}$ is particularly large form a curvaton web resembling a net of thick domain walls, strings, or global monopoles.}
\keywords{inflation, non-gaussianity, supersymmetry and cosmology} \arxivnumber{1012.0549}

\def\ba{\begin{array}}
\def\ea{\end{array}}
\def\be{\begin{equation}}
\def\ee{\end{equation}}
\def\bea{\begin{eqnarray}}
\def\eea{\end{eqnarray}}
\usepackage{amssymb}
\usepackage{graphicx}
\input epsf
\begin{document}
\maketitle

\section{Introduction}

One of the main reasons to introduce the curvaton scenario \cite%
{Linde:1996gt,Enqvist:2001zp,Lyth:2001nq,Moroi:2001ct} was to obtain a
realistic mechanism of generation of non-gaussian adiabatic perturbations of
metric \cite{Linde:1996gt,Lyth:2002my}. Since that time, many interesting
curvaton models were proposed. However, it would be nice to have a curvaton
model which would be as simple as the basic chaotic inflation scenario with
the potential $m^{2}\phi^{2}/2$ \cite{Linde:1983gd}. It would be good also
to find a natural implementation of this scenario in the context of
supergravity. This is the main goal of our work.

In this paper we will describe a broad family of models of chaotic inflation
in supergravity, which provide a natural realization of the curvaton theory.
We will calculate the non-gaussianity parameter $f_{\mathrm{NL}}$ for the
simplest versions of these models and show that this parameter takes
different values in different parts of the universe, in agreement with the
curvaton web scenario of Ref. \cite{Linde:2005yw}. If inflation is
sufficiently long, the average value of $f_{\mathrm{NL}}$ in this scenario
does not depend on the initial value of the curvaton field. We will also
show that under certain conditions the parameter $f_{\mathrm{NL}}$ can take
values in the observationally interesting range from $O(10)$ to $O(100)$.

\section{Curvaton scenario and chaotic inflation in supergravity}

\label{model}

For many years, it seemed very difficult to realize chaotic inflation in
supergravity. This problem was solved in \cite{Kawasaki:2000yn}. The authors
proposed a very simple model describing two fields, $S$ and $\Phi$, with the
superpotential 
\begin{equation}
W=m S\Phi,  \label{W3}
\end{equation}
and K{\"a}hler\, potential 
\begin{equation}
\mathcal{K} = S \bar S - \frac{1}{2}(\Phi -\bar\Phi)^2 .  \label{K3}
\end{equation}
Note that the K{\"a}hler\, potential does not depend on the phase of the
field $S$ and on the real part of the field $\Phi$. Therefore it will be
convenient for us to represent the fields $S$ and $\Phi$ as $S=\sigma\,
e^{i\theta}/\sqrt 2$ and $\Phi=(\phi+i\chi)/\sqrt 2$. The field $\phi$ plays
the role of the inflaton field, with the quadratic potential, as in the
simplest version of the chaotic inflation scenario \cite{Linde:1983gd}: 
\begin{equation}  \label{inflpot3}
V(\phi) = 3H^{2}= {\frac{m^2}{2}} \phi^2\ ,
\end{equation}
where $H$ is the Hubble constant during inflation. Near the inflationary
trajectory with $S = 0$, the mass squared of the imaginary part of the field 
$\Phi$ is $m^2_{\chi} = 6H^2 + m^2.$ Thus during inflation $m^2_{\chi} > 6
H^{2}$, and therefore the imaginary part of the field $\Phi$ is stabilized
at $\mathrm{Im}\, \Phi = 0$. No perturbations of this field are generated.

Both components of the field $S$ may remain light during inflation, and
therefore inflationary perturbations of these fields can be generated \cite%
{Davis:2008fv}. Since the potential does not depend on the field $\theta$,
we will ignore fluctuations of this field in our study of the curvaton
perturbations. The potential of the fields $\phi$, $\sigma$ at $\chi = 0$ is 
\begin{equation}  \label{corrected}
V(\phi,\sigma) = {\frac{m^{2}}{2}} e^{\sigma^2/2}\left[\phi^{2}+ \sigma^{2} +%
{\frac{\phi^{2}}{4}}\sigma^{2}(\sigma^{2}-2)\right] \ .
\end{equation}
For $\sigma \ll 1$ one has 
\begin{eqnarray}  \label{corrected1}
V(\phi,\sigma) = {\frac{m^{2}\phi^{2}}{2}}+ {\frac{m^{2}\sigma^{2}}{2}}+ {%
\frac{m^{2}\phi^{2}\sigma^{4}}{16}} \ .
\end{eqnarray}
The effective mass squared of the field $\sigma$ at $\sigma \ll 1$ is given
by 
\begin{equation}  \label{corrected2}
m^2_{\sigma} =V_{\sigma\sigma} = m^2 +{\frac{3}{4}} m^{2}\phi^{2}\sigma^{2}
= m^2 +{\frac{9}{2}}H^{2}\sigma^{2} \ ,
\end{equation}
where $V_{\sigma\sigma}$ means second partial derivative of $V$ with respect
to $\sigma$. One can easily see that $m^2_{\sigma} = m^2$ for $\phi\sigma
\ll 1$. During inflation $m^{2} \ll H^{2}$, and therefore inflationary
perturbations of the field $\sigma$ can be generated. At $\phi\sigma \gtrsim
1$, the effective mass squared of the field $\sigma$ is dominated by the
term ${\frac{3}{4}} m^{2}\phi^{2}\sigma^{2} = {\frac{9}{2}}H^{2}\sigma^{2} >
m^{2}$. For $\sigma \ll 1$ one still has $m^2_{\sigma} \ll H^{2}$, so the
perturbations of the field $\sigma$ are generated in this regime as well.
However, at $\sigma \gtrsim 1$ the potential becomes exponentially steep,
and $m^2_{\sigma} \gg H^{2}$. Therefore inflationary fluctuations of this
field are generated only for $\sigma \lesssim 1$. This is a very important advantage of the curvaton scenario in supergravity: The steepness of the curvaton potential at $\sigma \gtrsim 1$ protects us from extremely large perturbations of the curvaton field which otherwise could be produced during eternal inflation in this scenario \cite{Linde:2005yw,Eternal}.

If one does not take into account the curvaton fluctuations in this scenario
and study only the usual inflaton fluctuations \cite{Pert}, then the COBE
normalization requires $m \sim 6 \times 10^{-6}$, in the system of units $%
M_{p}=1$ \cite{Linde:1983gd,Linde:2007fr,book,mukhbook}. Thus, the mass of
the inflaton field must be somewhat smaller than $6 \times 10^{-6}$ if we
want to add the curvaton fluctuations to the inflaton fluctuations.

Recently the supergravity model described above was substantially
generalized in \cite{Kallosh:2010ug,Kallosh:2010xz}. The generalized
scenario describes a theory with a superpotential 
\begin{equation}
W=S f(\Phi),  \label{W3a}
\end{equation}
where $f(\Phi)$ is an arbitrary real holomorphic function. The K{\"a}hler\,
potential in this class of models may take several different functional
forms, e.g. 
\begin{equation}
\mathcal{K} = S \bar S - \frac{1}{2}(\Phi -\bar\Phi)^2 - {\frac{\alpha}{12}}
(S\bar S)^2 .  \label{K2}
\end{equation}
In this theory, the inflaton potential is given by 
\begin{equation}
V(\phi) = f^{2}(\phi/\sqrt 2).  \label{K2a}
\end{equation}
and the mass of the field $\sigma$ is 
\begin{equation}
m^2_{\sigma} = \alpha H^2 + (f^{\prime }(\phi/\sqrt 2))^{2} .
\end{equation}
In this class of models, one can implement chaotic inflation in
supergravity, with an \textit{arbitrary} shape of the inflaton potential $%
V(\phi)$. In all of these models one has $H^{2} = f^{2}(\phi/\sqrt 2)/3$.
The term $(f^{\prime }(\phi/\sqrt 2))^{2}$ is equal to $3H^{2}\epsilon$,
where $\epsilon \ll 1$ is the slow roll parameter. For $\alpha \gtrsim 1$
one has $m^2_{\sigma} \gtrsim H^{2}$. In this case no curvaton perturbations
are produced, so all standard predictions of the single-field inflaton
scenario remain intact.

On the other hand, in models with $\alpha \ll 1$ one has $m^2_{\sigma} \ll
H^{2}$ during inflation, which means that quantum fluctuations of the field $%
\sigma$ are generated during inflation \cite{Kallosh:2010ug,Kallosh:2010xz}.

Thus we have a broad class of models of chaotic inflation where the curvaton
scenario can be realized. One can further generalize this scenario by adding
terms $\sim S^{3}$ to the superpotential, and by using other versions of the
K{\"a}hler\, potential, as long as the K{\"a}hler\, potential has certain
properties described in \cite{Kallosh:2010ug,Kallosh:2010xz}. The
requirements which are necessary for the existence of the light curvaton
fields in this class of models can be formulated in an invariant way in
terms of the curvature of the K{\"a}hler\, geometry. In particular, the
parameter $\alpha$ is related to the curvature of the K{\"a}hler\, manifold 
\cite{Kallosh:2010xz}. The field $\sigma$ itself has an interesting
interpretation from the point of view of supergravity: it is the scalar
component $\sigma$ of the goldstino multiplet. Because of the generality and
simplicity of this scenario and because of its supergravity origin, one may
call it the \textit{supercurvaton scenario}.

In this paper we will concentrate on the simplest model (\ref{W3}), (\ref{K3}%
), but with an additional term $- {\frac{\alpha}{12}} (S\bar S)^2$ in the K{%
\"a}hler\, potential, as in Eq. (\ref{K2}). In this model the curvaton mass
squared along the inflationary trajectory with $\sigma = 0$ is given by 
\begin{equation}  \label{mass}
m^2_{\sigma} = m^{2} +\alpha H^2 \ ,
\end{equation}
and in a more general case $0< \sigma \ll 1$ the effective mass squared of
the field $\sigma$ is 
\begin{equation}  \label{mass2}
m^2_{\sigma} = m^{2} +\alpha H^2 +{\frac{9}{2}}H^{2}\sigma^{2} = m^{2} +{%
\frac{\alpha}{6}} m^2\phi^{2} +{\frac{3}{4}} m^{2}\phi^{2}\sigma^{2} \ .
\end{equation}


\section{Curvaton perturbations and non-gaussianity}

During inflation, the curvaton perturbations are produced. An average
amplitude of perturbations produced during each Hubble time $H^{{-1}}$ is
given by $\delta\sigma = {\frac{H}{2\pi}}$. Then these fluctuations are
stretched, overlap with each other, and eventually produce a classical
curvaton field $\sigma$ which look relatively homogeneous in the observable
part of the universe, but may take different values in other parts of the
universe \cite{Linde:2005yw}. The amplitude of the perturbations of density
of the curvaton field with a quadratic potential is given by ${%
\delta\rho_{\sigma}/ \rho_\sigma} \sim {2\delta\sigma/ \sigma}$. However,
the total energy density of matter at the moment when the curvaton field
decays may be greater than the energy of the classical field $\sigma$. This
may happen, for example, if the decay of the inflaton field during reheating
produces many curvaton particles \cite{Linde:2005yw}. Therefore the relative
perturbation of density will be given by 
\begin{equation}  \label{sigma}
{\frac{\delta\rho_{\sigma}}{\rho}} \sim {\frac{2r\delta\sigma}{\sigma}} \ ,
\end{equation}
where $r = \rho_{\sigma}/\rho$ at the time of the curvaton decay. According
to \cite{Lyth:2002my}, these perturbations will match the COBE normalization
of the spectrum for 
\begin{equation}  \label{COBE}
r{\frac{\delta\sigma}{\sigma}} \sim 7 \times 10^{-5} \ .
\end{equation}
These perturbations are non-gaussian, with the amplitude of local
non-gaussianity given by \cite{Lyth:2002my} 
\begin{equation}  \label{fnl}
f_{\mathrm{NL}} = {\frac{5}{4 r}} \ .
\end{equation}

Our goal will be to find a typical value of $\sigma$ in some of the simplest
supergravity models described above, calculate $\delta\sigma$, find the
value of $r$ required to satisfy Eq. (\ref{COBE}), and finally determine $f_{%
\mathrm{NL}}$. The most complicated part of this program is finding a
typical value of $\sigma$.

\section{Stochastic approach}

\label{stoch}

We will begin our study with investigation of the behavior of the
distribution of the fluctuations of the curvaton field $\sigma$ with a
simple quadratic potential $m^{2}_{\sigma}\sigma^{2}/2$. This approach will
allow us to describe the case when $m^2_{\sigma} = m^{2} +\alpha H^2$, but
not the more general situation when $m^{2}_{\sigma}$ depends on $\sigma$ as
in (\ref{mass2}), which will be discussed separately.

During inflation, the long-wavelength distribution of this field generated
at the early stages of inflation behaves as a nearly homogeneous classical
field, which satisfies the equation%
\begin{equation}
3H\dot{\sigma}+V_{\sigma }=0\ .  \label{a1}
\end{equation}%
or, equivalently, 
\begin{equation}
{\frac{d\sigma ^{2}}{dt}}=-{\frac{2V_{\sigma }\,\sigma }{3H}}\ .
\label{a3ax011}
\end{equation}%
However, each time interval $H^{{-1}}$ new fluctuations of the scalar field
are generated, with an average amplitude squared 
\begin{equation}
{\langle \delta \sigma ^{2}\rangle }={\frac{H^{2}}{4\pi ^{2}}}\ .
\label{a11}
\end{equation}%
The wavelength of these fluctuations is rapidly stretched by inflation. This
effect increases the average value of the squared of the classical field $%
\sigma $ in a process similar to the Brownian motion. As a result, the
square of the field $\sigma $ at any given point with an account taken of
inflationary fluctuations changes, in average, with the speed which differs
from the predictions of the classical equation of motion by ${\frac{H^{3}}{%
4\pi ^{2}}}$: 
\begin{equation}
{\frac{d\sigma ^{2}}{dt}}=-{\frac{2V_{\sigma }\,\sigma }{3H}}+{\frac{H^{3}}{%
4\pi ^{2}}}\ .  \label{a3ax0}
\end{equation}%
Using $3H\dot{\phi}=-V_{\phi }$, one can rewrite this equation as 
\begin{equation}
{\frac{d\sigma ^{2}}{d\phi }}={\frac{2V_{\sigma }\,\sigma }{V_{\phi }}}-{%
\frac{V^{2}}{12\pi ^{2}V_{\phi }}}\ .  \label{a3ax}
\end{equation}%
It's solution with the initial condition $\sigma (\phi _{i})=0$ for $%
m_{\sigma }^{2}=m^{2}+\alpha H^{2}$ is given by 
\begin{equation}
\sigma ^{2}\left( \phi \right) =\frac{1}{12\pi ^{2}}\int\limits_{\phi
}^{\phi _{i}}\frac{V^{2}(\tilde{\phi})}{V^{\prime }(\tilde{\phi})}\exp
\left( -2\int\limits_{\phi }^{\tilde{\phi}}\frac{m_{\sigma }^{2}}{V^{\prime
}(\bar{\phi})}d\bar{\phi}\right) d\tilde{\phi}\ ,  \label{a6}
\end{equation}%
where $\phi _{i}$ is the initial value of the inflaton field. 

If inflation continued for much longer time than 60 e-foldings, as we will
assume in this paper, the main contribution to $\sigma$ is given by
perturbations produced at the very early stages of inflation. Such
fluctuations look almost absolutely homogeneous on the scale of the
observable part of our universe, so our calculations give us a typical value
of the classical field $\sigma$ inside the observable part of our universe.
However, in different parts of the universe, the field $\sigma$ may be
significantly smaller or greater than its ``typical'' value calculated
above. As a result, the amplitude of the curvaton perturbations is not a
constant, but it varies in space \cite{Linde:2005yw}. The same is true for
the degree of non-gaussianity $f_{\mathrm{NL}}$, see Section \ref{fnlsect}.

Therefore, to be precise, one should distinguish between the average amplitude of the field $\sigma$ calculated above, when the averaging it taken all over the universe, and the local value of the field $\sigma$ in each horizon-size part of the universe. We will make this distinction in Section  \ref{fnlsect}, where we will make a slight change of notation and call the value of the curvaton field averaged over the whole universe $\bar\sigma$, reserving the letter $\sigma$ for the average value of the curvaton field in the horizon-size part of the universe. However, in the main part of our paper we will not distinguish between $\sigma$ and $\bar\sigma$. This means, in particular, that when we will calculate $f_{\mathrm{NL}}(\sigma)$, our results will in fact describe the value of this parameter for $\sigma = \bar\sigma$, i.e. the value of $f_{\mathrm{NL}}$ for an average value of $\sigma$, all over the universe. In Section \ref{fnlsect} we will show, however, that the value of $f_{\mathrm{NL}}$ for an average value of $\sigma$ can be significantly different from the average value of $f_{\mathrm{NL}}$; the order of averaging in certain cases can be very important. One should take this effect into account when making predictions of the non-gaussianity in each particular curvaton scenario.

We should note also that in general the curvaton field may not be equal to zero
at the beginning of chaotic inflation, so one may also consider a
possibility that initially $\langle \sigma ^{2}\rangle (\phi _{i})$ was very
large. In this respect, the supergravity model which we are going to study
provides an important simplification: The curvaton field initially cannot be
much larger than $O(1)$ because of the exponential steepness of the
potential at $\sigma \gtrsim 1$. Also, the effective mass term ${\frac{3}{4}}%
m^{2}\phi ^{2}\sigma ^{2}$ in the supergravity potential (\ref{corrected}),
rapidly reduces the initial value of the field $\sigma $, thus making
quantum fluctuations generated during inflation more important than the
initial value of the classical field $\sigma $. We will study these issues
in the rest of the paper, starting from the simple toy model with $m_{\sigma
}^{2}=m^{2}$ and ending up with the model with $m_{\sigma }^{2}=m^{2}+{\frac{%
\alpha }{6}}m^{2}\phi ^{2}+{\frac{3}{4}}m^{2}\phi ^{2}\sigma ^{2}$. As we
will see, in all these cases the final result does not depend on the initial
distribution of the curvaton field if inflation lasts long enough.

\section{Nongaussianity in various regimes}

\boldmath
\subsection{ A toy model with { $m_{\protect\sigma}^{2}=m^{2}$}}
\unboldmath

\label{constant} In this subsection we will study the distribution of the
curvaton field with the mass $m^{2}_{\sigma} = m^{2}$ during inflation
driven by the massive inflaton field with potential $V=\frac{1}{2}m^{2}\phi
^{2}.$ 
In this case Eq. (\ref{a6}) implies that the classical scalar field $\sigma $
which is nearly homogeneous on the scale of the horizon has a typical
amplitude 
\begin{equation}
\sigma (\phi)={\frac{m\,\phi\, \phi _{i}}{4\pi \sqrt{6}}} \ .  \label{8aaaa}
\end{equation}

Meanwhile the amplitude of fluctuations of $\sigma $ generated at that time
is 
\begin{equation}
\delta \sigma \sim {\frac{H}{2\pi }}={\frac{m\phi }{2\pi \sqrt{6}}}\ .
\end{equation}%
During the subsequent evolution of the universe, $\sigma $ and $\delta
\sigma $ both decrease in the same way, and therefore at the end of
inflation the curvaton perturbations have flat spectrum with the amplitude 
\begin{equation}
{\frac{\delta \sigma }{\sigma }}={\frac{2}{\phi _{i}}}.
\end{equation}%
As we mentioned above the amplitude of the perturbations must be normalized
as 
\begin{equation}
r{\frac{\delta \sigma }{\sigma }}\simeq {\frac{2r}{\phi _{i}}}\sim 7\times
10^{{-5}}\ .  \label{a14bc}
\end{equation}%
and hence 
\begin{equation}
f_{\mathrm{NL}}={\frac{5}{4r}}\sim {\frac{3.5\times 10^{4}}{\phi _{i}}}.
\label{fnl2}
\end{equation}%
This means that the degree of nongaussianity depends on the initial value of
the inflaton field. Unless this field is very large, $f_{\mathrm{NL}}$ may
be extremely large.

However, in supergravity models which we study in this paper the approach
developed above is valid only if inflation was short enough, that is, $\phi
_{i}\ll m^{-1/3}$ and for large values of $\phi _{i}$ one cannot ignore the
supergravity correction to the mass in Eq. (\ref{corrected2}).

\boldmath
\subsection{{ $m^2_{\protect\sigma} = m^{2} +{\frac{9}{2}}H^{2}%
\protect\sigma^{2}$}}
\label{total}
\unboldmath

In the previous section we made a simplifying assumption that the curvaton
mass does not depend on $\sigma $, which allowed us to use Eq. (\ref{a6}).
However, as one can see from (\ref{corrected}), in supergravity model (\ref%
{W3}), (\ref{K3}) the curvaton mass does depend on $\sigma $ in a rather
complicated way. The leading correction to the curvaton mass squared $m^{2}$
is given by ${\frac{3}{4}}m^{2}\phi ^{2}\sigma ^{2}={\frac{9}{2}}H^{2}\sigma
^{2}$ and it becomes dominant for $\phi \sigma \gtrsim 1$.

To find out how it will change the final result one has to solve Eq. (\ref%
{a3ax}) for 
\begin{equation}\label{RRR}
V={\frac{m^{2}\phi ^{2}}{2}}+{\frac{m^{2}\sigma ^{2}}{2}}+{\frac{m^{2}\phi
^{2}\sigma ^{4}}{16}}\ ,
\end{equation}%
which takes in this case the following form 
\begin{equation}
y^{\prime }={\frac{y}{x}}+{\frac{y^{2}}{4}}-bx\ ,  \label{airy}
\end{equation}%
where $x=\phi ^{2}$, $y=\sigma ^{2}$, and $b={\frac{m^{2}}{96\pi ^{2}}}$.

The general solution of this equation can be expressed in terms of Airy
functions, 
\begin{equation}
y(x)=-(2b)^{2/3}x\,{\frac{\mathrm{Ai}(z)-c\,\mathrm{Bi}(z)}{\mathrm{%
Ai^{\prime }}(z)-c\,\mathrm{Bi^{\prime }}(z)}}\ ,  \label{a6xx}
\end{equation}%
where $z=2^{-2/3}b^{1/3}x$ and $c$ is a small constant which should be
chosen in such a way that $y(x_{i})=0$.

Suppose first that the initial value of the field $\phi $ is much higher
than $m^{{-1/3}}$, i.e. $z\gg 1$. One can check that in this case one should
take $c\ll 1$ to have $y(x_{i})=0$. Inflation ends at $\phi \sim 1$, when $%
z\ll 1$. In this limit, all functions are $O(1)$. Therefore the functions $%
\mathrm{Bi}(z)$ drop out from the final expression because of the small
coefficient $c$, $\mathrm{Ai}(z)\approx \mathrm{Ai}(0)=3^{{-2/3}}\,\Gamma ^{{%
-1}}(2/3)$, $\mathrm{Ai}^{\prime }(z)\approx \mathrm{Ai}^{\prime
-1/3}\,\Gamma ^{{-1}}(1/3)$. As a result, 
\begin{equation}
y(x)\approx -{\frac{(2b)^{2/3}\,x\,\,\Gamma (1/3)}{3^{{1/3}}\,\Gamma (2/3)}}%
\ .  \label{a6xxx}
\end{equation}%
Expressing everything in terms of the original fields $\phi $ and $\sigma $,
we find 
\begin{equation}
\sigma (\phi )\approx -{\frac{m^{2/3}\,\phi \,\,}{2^{{4/3}}\sqrt{3}\pi
^{2/3}\,}}{\frac{\Gamma (1/3)}{\Gamma (2/3)}}\approx 0.15\ m^{2/3}\,\phi \ .
\label{a6xxxx}
\end{equation}%
This yields 
\begin{equation}\label{alpha0}
{\frac{\delta \sigma }{\sigma }}\sim 0.4\,m^{1/3}.
\end{equation}%
The COBE normalization requires $rm^{1/3}\sim 7\times 10^{-5}$. Therefore,
for $m\sim 10^{{-7}}$ we have $r\sim 0.04$ and $f_{\mathrm{NL}}\sim 30$.

If, on the other hand, the initial value of field $\phi $ is much smaller
than $m^{{-1/3}}$, then the final result looses it's universality and become
sensitive to $\phi _{i}$. In this case, one can either use the analytical
solution above, with different initial conditions, or simply use the results
of the previous section (one can see that in this case $\phi \sigma \ll 1$,
and hence the  results of Section \ref{constant} are valid).

Note that in our calculation of $f_{{\rm NL}}$ we used equation (\ref{fnl}), which was obtained in  \cite{Lyth:2002my} under the assumption that the curvaton potential is purely quadratic. Meanwhile in our case the curvaton potential contains the quadratic term ${\frac{m^{2}\sigma ^{2}}{2}}$ as well as the quartic term ${\frac{m^{2}\phi
^{2}\sigma ^{4}}{16}}$, see (\ref{RRR}). This could lead to some corrections to equation (\ref{fnl})  \cite{Sasaki:2006kq}. Fortunately, one can show that during the last 60 e-foldings of inflation in our model the quartic term is vanishingly small as compared to the quadratic term. That is why one can use the simple equation (\ref{fnl}) for the calculation of $f_{{\rm NL}}$.

\boldmath
\subsection{ {$m^{2}_{\protect\sigma} =m^{2}+\protect\alpha H^{2},$%
~ $\protect\alpha >0$}}

\label{alpha}
\unboldmath

Now we will consider the case when the mass of the curvaton field is given
by 
\begin{equation}
m_{\sigma }^{2}=\alpha H^{2}+m^{2}=m^{2}\left( {\frac{\alpha \,\phi ^{2}}{6}}%
+1\right) ,
\end{equation}%
where we have ignored the correction ${\frac{3}{4}}m^{2}\phi ^{2}\sigma ^{2}$, which will be taken into account in Section \ref{full}.

To study this case we consider separately the evolution of perturbations at $%
\alpha \phi ^{2}/6>1$ and $\alpha \phi ^{2}/6<1$ assuming that during last
60 e-folds of inflation the condition $\alpha \phi ^{2}/6<1$ is satisfied,
which means that $\alpha \lesssim 1/40$. Thus during last 60 e-folds, $%
m_{\sigma }^{2}\approx m^{2}$, and hence one can use the results of Section %
\ref{stoch}.

Substituting $m^{2}=\alpha H^{2}=\frac{\alpha }{3}V$ in (\ref{a6}), we obtain%
\begin{equation}
\sigma ^{2}\left( \phi \right) =\frac{1}{12\pi ^{2}}\int\limits_{\phi
}^{\phi _{i}}\frac{V^{2}(\tilde{\phi})}{V^{\prime }}\exp \left( -\frac{%
2\alpha }{3}\int\limits_{\phi }^{\tilde{\phi}}\frac{V}{V^{\prime }}d\bar{\phi%
}\right) \,d\tilde{\phi}  \label{a12}
\end{equation}%
For the case of the power-law potential $V$ the integral in (\ref{a12}) can
be calculated exactly. In particular, for $V=\frac{1}{2}m^{2}\phi ^{2}$ and $%
\alpha \gg \phi _{i}^{-2}\sim m\gg \phi ^{2}$ one obtains 
\begin{equation}
\sigma ^{2}\left( \phi \right) =\frac{m^{2}}{16\pi ^{2}\alpha }\left( \phi
^{2}+\frac{6}{\alpha }\right) .  \label{a13}
\end{equation}%
Note that that this result does not depend on the initial value of the
inflaton field. At the end of the first stage of inflation when $\alpha \phi
_{1}^{2}/6=1$, both terms in the brackets are equal to each other and the
averaged value of $\sigma $ at that time is about 
\begin{equation}
\sigma (\phi _{1})\simeq \frac{\sqrt{3}\,m}{2\pi \alpha }\ ,  \label{a14}
\end{equation}%
while the amplitude of the perturbations of field $\sigma $ is 
\begin{equation}
\delta \sigma (\phi _{1})\simeq \frac{m}{2\pi \sqrt{\alpha }}\ .
\label{a14a}
\end{equation}%
The CMB normalization of the amplitude of the perturbations thus requires 
\begin{equation}
r{\frac{\delta \sigma }{\sigma }}\simeq r\sqrt{\frac{\alpha }{3}}\sim
7\times 10^{{-5}}\ .  \label{a14b}
\end{equation}%
Note that in this case the amplitude of the curvaton perturbations does not
depend on the inflaton mass $m$.

Taking $\alpha =10^{-4}$ we find that $r\approx 10^{{-2}}$ and hence $f_{%
\mathrm{NL}}\sim 10^{2}$. Meanwhile for $\alpha =10^{-2}$ we should have $%
r\approx 10^{{-3}}$, which gives $f_{\mathrm{NL}}\sim 10^{3}$.

One may wonder what is the origin of such an incredible sensitivity of the
results to the choice of the parameter $\alpha$. The answer is that this
parameter makes the mass of the curvaton field much greater than the mass of
the inflaton field at the early stages of inflation. As a result, the
distribution of the field $\sigma$ shrinks fast while the field $\phi$ rolls
down.

In the calculations above we have ignored the supergravity correction to the
curvaton mass squared: ${\frac{3}{4}}m^{2}\phi ^{2}\sigma ^{2}={\frac{9}{2}}%
H^{2}\sigma ^{2}.$ As we will show in the next section this correction can
be ignored only if $\alpha \gg 10^{-1}m^{{2/3}}$ and hence the results of
this section are applicable only in this case.

\boldmath
\subsection{{$m^2_{\protect\sigma} = m^{2} +{\protect\alpha}
H^{2} +{\frac{9}{2}} H^{2}\protect\sigma^{2}$, $\protect\alpha > 0$}}
\label{full} 
\unboldmath

Now we will study the curvaton perturbations in the theory with
the general potential

\begin{equation}
V={\frac{m^{2}\phi ^{2}}{2}}+{\frac{m^{2}\sigma ^{2}}{2}}+{\frac{m^{2}\phi
^{2}\sigma ^{4}}{16}}+\alpha {\frac{m^{2}\phi ^{2}\sigma ^{2}}{6}}\ ,
\end{equation}%
which corresponds to the curvaton mass (\ref{mass2}). For $\sigma ^{2}\ll 1$
equation (\ref{a3ax}) becomes 
\begin{equation}
{\frac{dy}{dx}}={\frac{y}{x}}+{\frac{\alpha y}{3}}+{\frac{y^{2}}{4}}-bx\ ,
\label{y11}
\end{equation}%
where $x=\phi ^{2}$, $y=\sigma ^{2}$, $b={\frac{m^{2}}{96\pi ^{2}}}$. In
this case, unlike to Eq. (\ref{airy}), there is no exact analytical
solution. Nevertheless one can investigate the solutions of this equation
using phase diagram method. If inflation lasts long enough, all solutions,
independently of the initial conditions, converge at a certain attractor
trajectory in the phase space $(y,x)$, or, equivalently in the space $%
(\sigma ,\phi )$, see Fig. 1. For large $\phi $, this attractor trajectory
is given in the leading order by the solution of the algebraic equation: 
\begin{equation}
{\frac{\tilde{y}^{2}}{4}}+\tilde{y}\left( {\frac{1}{x}}+{\frac{\alpha }{3}}%
\right) -bx=0,  \label{algebr}
\end{equation}%
which is 
\begin{equation}\label{funct}
\tilde{y}(x)=-2\left( {\frac{1}{x}}+{\frac{\alpha }{3}}\right) +2\sqrt{%
\left( {\frac{1}{x}}+{\frac{\alpha }{3}}\right) ^{2}+bx}.
\end{equation}

\begin{figure}[h!]
\centering
\vskip -0.7cm \includegraphics[scale=0.4]{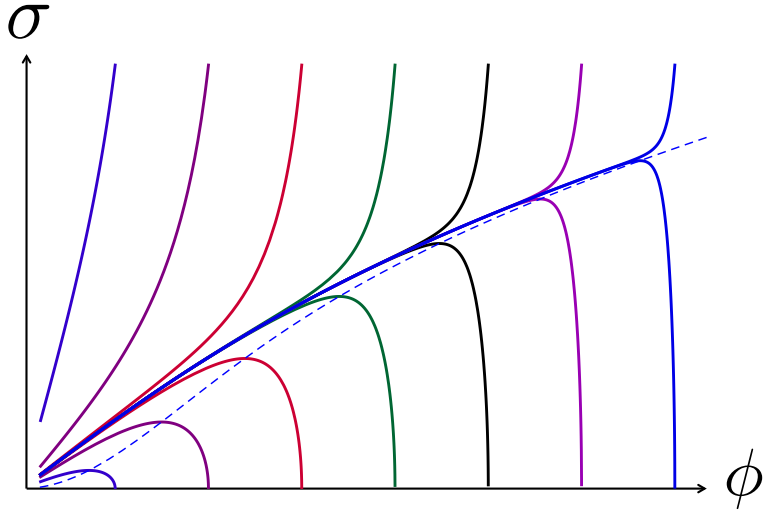}
\caption{Behavior of the average value of the curvaton field $\protect\sigma 
$ as a function of the inflaton field $\protect\phi $, for various initial
conditions. As we see, all trajectories which start at the early stages of
inflation (large field $\protect\phi $) converge to the same attractor
solution. We follow it until the field $\protect\phi $ becomes $O(1)$ and
inflation ends. At large $\protect\phi $, this solution is very close to the square root of the
function (\protect\ref{funct}), which is shown by
the blue dashed line.}
\label{WMAP2}
\end{figure}

The existence of the attractor solution implies that if inflation is long
enough, the final results do not depend on the choice of initial conditions
for the curvaton field. We have also found above that in the limit $\alpha
\rightarrow 0$ one should get the asymptotic solution (\ref{a6xxxx}),
whereas for large $\alpha $ the asymptotic solution is given by (\ref{a13}),
(\ref{a14}). One may wonder how large should $\alpha $ become for the switch
between these asymptotic regimes?

To answer this question, let us use the variables: 
\begin{equation}
x=z\,b^{-1/3}\ ,\qquad y=u\,b^{{1/3}}\ ,\qquad \alpha =\gamma \,b^{{1/3}}\ ,
\end{equation}%
in terms of which equation (\ref{y11}) becomes 
\begin{equation}
{\frac{du}{dz}}={\frac{u}{z}}+{\frac{\gamma u}{3}}+{\frac{u^{2}}{4}}-z\ .
\label{y12}
\end{equation}%
%
After rewriting equation (\ref{y11}) in this form it becomes clear that the
behavior of the solutions is controlled by a single parameter $\gamma
=\alpha \,b^{{-1/3}}\sim 10\,\alpha \,m^{{-2/3}}$. One can easily understand
that the two asymptotic regimes discussed above corresponds to $\gamma \ll 1$
and $\gamma \gg 1$. One can confirm this conclusion by direct numerical
calculations.

This means that the results obtained in Section \ref{total} are valid for $%
\alpha \ll 10^{-1}m^{{2/3}}$. Meanwhile in the opposite limit $\alpha \gg
10^{-1}m^{{2/3}}$ one should use the results of Section \ref{alpha}. To give
a particular example, let us take $m\sim 10^{-6}$. In this case one can use
the results of Section \ref{total} for $\alpha \ll 10^{-5}$, whereas for $%
\alpha \gg 10^{-5}$ one should use the results of Section \ref{alpha}.

\boldmath
\subsection{{$m^{2}_{\protect\sigma} =m^{2}+ \protect\alpha %
H^{2}+{\frac{9}{2}} H^{2}\protect\sigma^{2}$,~ $\protect\alpha < 0$}}

\label{alpha2}
\unboldmath

Finally, we will study the case $\alpha <0$. At first glance in this model
the mass squared of the curvaton field at large $H^{2}$ and $\sigma =0$ is
negative, and therefore one expects a tachyonic instability. However,
similar to the model considered in the previous section, one can show that
for $|\alpha |\ll 10^{-1}m^{{2/3}}$ the effect related to the negative mass
squared contribution $\alpha H^{2}$ is subdominant and can be ignored. In
this case the results obtained in previous Section \ref{total} are
applicable.

For $10^{-1}m^{{2/3}}\ll |\alpha |\ll 1$, the tachyonic instability leads to
spontaneous symmetry breaking controlled by the supergravity correction ${%
\frac{9}{2}}H^{2}\sigma ^{2}={\frac{3}{4}}m^{2}\phi ^{2}\sigma ^{2}$ to the
curvaton mass squared. Indeed, one can show that the minimum of the
supergravity potential for the curvaton field, in the regime with $|\alpha
|,\sigma \ll 1$, can be found from the following equation: 
\begin{equation}
\sigma ^{2}={\frac{2|\alpha |}{3}}+{\frac{4}{\phi ^{2}}}\ .
\end{equation}%
Therefore at large $\phi $ and $\alpha <0$ the potential has a minimum at 
\begin{equation}
\sigma ^{2}={\frac{2|\alpha |}{3}}\ .
\end{equation}%
This means that during inflation the field $\sigma $ falls towards this
minimum, and its distribution become centered not at $\sigma =0$ but at $%
\sigma =\sqrt{\frac{2|\alpha |}{3}}$. As for the height of the potential
along the trajectory with $\sigma =\sqrt{\frac{2|\alpha |}{3}}$, for small $%
|\alpha |$ it remains approximately given by $m^{2}\phi ^{2}/2$.

When the field $\phi ^{2}$ becomes smaller than $6/|\alpha |$, the minimum
of the potential shifts towards $\sigma =0$, and the curvaton mass squared
becomes equal to $m^{2}$. However, this does not mean that the distribution
of the field $\sigma $ instantly follows the position of the minimum. Since
the mass of the field $\sigma $ at that time is much smaller than $H$, the
field $\sigma $ will move towards $\sigma =0$ very slowly, decreasing at the
same rate as the amplitude of perturbations $\delta \sigma $. As before, we
are assuming that $|\alpha |\lesssim 1/40$, and therefore the curvaton mass
squared is given by $m^{2}$ during the last 60 e-folds of inflation. This
leads to the following result for the perturbations: 
\begin{equation}
{\frac{\delta \sigma }{\sigma }}\sim {\frac{m\sqrt{3}}{2\sqrt{2}\pi
\,|\alpha |}}\ .
\end{equation}%
For $m\sim 10^{-7}$ and $\alpha \sim -10^{{-4}}$ a proper amplitude of
perturbations corresponds to $r\approx 1/3$ and, hence, $f_{\mathrm{NL}}=O(3)
$. However, one can easily increase $f_{\mathrm{NL}}$ by increasing $m$
and/or decreasing $|\alpha |$. For example, taking $m\sim 10^{-7}$ and $%
\alpha \sim -10^{{-5}}$ gives $f_{\mathrm{NL}}\sim 30$.

\section{Non-gaussianity and the curvaton web}

\label{fnlsect} In the previous sections we have evaluated the average value
of the curvaton field at the last stages of inflation, and calculated the
parameter $f_{\mathrm{NL}}$ describing local non-gaussianity. However, we
should remember that when we were talking about the classical homogeneous
curvaton field $\sigma $, we had in mind the long-wavelength perturbations
which look homogeneous on the scales corresponding to the present observable
part of the universe. In reality this classical field in our model is a
random variable with the expectation value $\bar{\sigma}=\sqrt{\left\langle
\delta \sigma ^{2}\right\rangle }$ obtained by summing up the contributions
of all long wavelength fluctuations (larger that the present horizon)
generated on inflation. All calculations above were performed taking $\sigma $
to be equal $\bar{\sigma}.$ However, because $\sigma $ is a random gaussian
variable it take different values in different parts of the universe of the
size of our horizon \cite{Linde:2005yw}. 

To evaluate the observational implications of this fact, let us try to understand how the amplitude of perturbations of metric and the local value of $f_{\rm NL}$ depend on the local value of $\sigma$. For simplicity, we will assume that the standard inflaton perturbations are very small, so that we can ignore them in our investigation. This can be achieved by considering a model with $m \ll 6\times 10^{{-6}}$. We will also assume that the curvaton field density at the moment of the curvaton decay is much smaller than the total density, i.e. $r \ll 1$. In this case, the change of $\sigma$ does not affect the total density $\rho$, but it does affect $\delta\rho(\sigma)$, which is proportional to $\sigma$. This means that the amplitude of perturbations of metric produced by fluctuations of the curvaton field will be proportional to ${\sigma\over\bar\sigma}$:
\be\label{u}
{\delta\rho(\sigma)\over\rho}  = {\delta\rho(\bar\sigma)\over\rho} \cdot {\sigma\over\bar\sigma} \ .
\ee
Meanwhile the local value of $f_{\rm NL}$ is inversely proportional to $r = {\rho(\sigma)\over\rho}$. For small $\sigma$, the value of $\rho(\sigma)$ is proportional to $\sigma^{2}$. Therefore
\be\label{v}
{f_{\rm NL}(\sigma)}  = {f_{\rm NL}(\bar\sigma)} \cdot {\bar\sigma^{2}\over\sigma^{2}} \ .
\ee

The probability that the curvaton
field will take some value much greater than $\bar{\sigma}$, is
exponentially small. However, the probability that $\sigma $ is
substantially smaller than $\bar{\sigma}$ can be rather large. To estimate
this probability we will make a simplifying assumption. Namely, we
assume that all values of the field $S=\sigma \,e^{i\theta }/\sqrt{2}$ with $%
|S|<\bar{\sigma}$ are equally probable, but the probability vanishes for $%
|S|>\bar{\sigma}$. In the maximal value of the curvaton field is $\bar{\sigma}\sqrt{2}$, the probability to find the field $\sigma$ in the interval $d\sigma$ from 0 to  $\bar{\sigma}\sqrt{2}$ is given by ${\sigma d\sigma\over\bar\sigma^{2}}$, 
and the average value of the curvaton field is $\bar\sigma$, as it should be. 


Now let us evaluate the average value of the amplitude of density perturbations, averaged over all possible values of $\sigma$:
\begin{equation}
\left\langle{\delta\rho(\sigma)\over\rho}\right\rangle \simeq {\delta\rho(\bar\sigma)\over\rho} \int\limits_{0}^{\bar{\sigma} \sqrt 2
}{\sigma\over\bar\sigma} \frac{\sigma d\sigma }{\bar{\sigma}^{2}}\simeq {2\sqrt 2\over 3}\, {\delta\rho(\bar\sigma)\over\rho}  \ . \label{1ppp}
\end{equation}%
Thus, the average amplitude of the curvaton perturbations almost exactly coincides with the amplitude of perturbations in the universe with an average curvaton field $\bar\sigma$.

However, the situation with $\left\langle f_{\mathrm{NL}}\right\rangle$ is quite different. Since $f_{\mathrm{%
NL}}$ is proportional to $\sigma ^{-2}$,  it expectation value over the whole
universe acquires a divergent contribution from the parts of the universe with small $\sigma$. Our calculations are valid only for fluctuations produced well before the last 60 e-folds of inflation, with a combined amplitude $\sigma$ above $O(H)$. Introducing the cut-off at $\sigma \sim H \sim 2\pi \delta\sigma$, we find 
\begin{equation}
\left\langle f_{\mathrm{NL}}\right\rangle \simeq f_{\mathrm{NL}}\left( \bar{\sigma}\right) \int\limits_{H}^{\bar{\sigma%
}\sqrt 2} \left( \frac{\bar{\sigma}}{%
\sigma }\right) ^{2}\frac{\sigma d\sigma }{\bar{\sigma}^{2}}\simeq f_{%
\mathrm{NL}}\left( \bar{\sigma}\right) \ln \left( \frac{\bar{\sigma}}{\sqrt 2\,\pi \delta\sigma}%
\right) .  \label{1aaa}
\end{equation}%

How significant is the effect discussed above? To give a particular numerical example, let us consider the case $\alpha = 0$. In this case, according to Eq. (\ref{alpha0}), one has
${\frac{\delta \sigma }{\bar\sigma }}\sim 0.4\,m^{1/3}.$ We found that for $m\sim 10^{-7}$ one has $f_{\rm NL}(\bar\sigma) \sim 30$. In this case Eq. (\ref{1aaa}) implies that $\left\langle f_{\mathrm{NL}}\right\rangle \sim 5\, f_{\rm NL}(\bar\sigma) \sim 150$.

Thus we deal with a significant effect of statistical amplification of non-gaussianity: Even though the fraction of the volume of the universe with $f_{\rm NL}(\sigma) \gg  f_{\rm NL}(\bar\sigma)$ is relatively small, the values of $f_{\rm NL}$ in those parts of the universe can be huge, so the expectation value of $ f_{\rm NL}$ can be much greater than the value of this parameter $ f_{\rm NL}(\bar\sigma)$ calculated in the previous sections of the paper.

This effect becomes even stronger in the models where the curvaton field is real (instead of being a radial part of a complex field). In such models
\begin{equation}
\left\langle f_{\mathrm{NL}}\right\rangle \simeq f_{\mathrm{NL}}\left( \bar{\sigma}\right) \int\limits_{H}^{2\bar{\sigma%
}} \left( \frac{\bar{\sigma}}{%
\sigma }\right) ^{2}\frac{d\sigma }{\bar{\sigma}}\simeq f_{%
\mathrm{NL}}\left( \bar{\sigma}\right)  \frac{\bar{\sigma}}{2\,\pi \delta\sigma} .  \label{1aaaq}
\end{equation}%

In the particular example discussed above, ${\frac{\delta \sigma }{\bar\sigma }}\sim 0.4\,m^{1/3}$ and $m\sim 10^{-7}$, this would lead to an enormously large amplification effect:   $\left\langle f_{\mathrm{NL}}\right\rangle \sim 10^{2}\, f_{\rm NL}(\bar\sigma) \sim 3000$.

Thus we see that in the curvaton scenario some fraction of the
universe can be in a state with the curvaton field $\sigma$ significantly smaller than its average value $\bar\sigma$. In such parts of the universe, the locally observed level of non-gaussianity will strongly exceed its value $f_{\rm NL}(\bar\sigma)$
calculated in the previous sections. This effect is so significant that the average value of the parameter $f_{\rm NL}$ can be much greater than the value $f_{\rm NL}$ in the part of the universes with an average value of the field $\sigma$. In other words, operations of averaging in this case are not commutative.

For a complete investigation of this effect one should also take into account the standard inflationary perturbations of metric. The curvaton perturbations are important only in the cases where the standard inflaton perturbations are suppressed. That is why we assumed that $m \ll 6\times 10^{{-6}}$. But the standard inflaton perturbations may dominate in the rare parts of the universe where  $\sigma \ll \bar\sigma$. In such cases one should perform a more detailed investigation of nongaussianity of perturbations produced by all sources.

This means that one should be very careful when formulating predictions for the nongaussianity parameter $f_{\rm NL}$ in the curvaton scenario, because the distribution of possible values of $f_{\mathrm{NL}}$ in the curvaton web can be extremely broad. 
Moreover, the existence of the anti-correlation between the amplitude of the perturbations of metric $\left({\delta\rho(\sigma)\over\rho}\right)^{2}$ and the non-gaussianity parameter $f_{\mathrm{NL}}$ for $r\ll 1$ (see equations (\ref{u}) and (\ref{v})) suggests that anthropic considerations may play a very important role in evaluation of the probability to live and make observations in parts of the curvaton web with different values of the non-gaussianity parameter  $f_{\mathrm{NL}}$ \cite{Linde:1996gt,Linde:2005yw,GarciaBellido:1993wn,Tegmark:1997in,Garriga:2005ee,Lyth:2006gd}. We hope to return to the discussion of this issue in a separate publication.

The difference between $\left\langle f_{\mathrm{NL}}\right\rangle$ and $f_{\rm NL}(\bar\sigma)$ clearly demonstrates that $f_{\rm NL}$ is not a perfect tool for the description of non-gaussianity.  As  shown in  \cite{Linde:2005yw}, the distribution of the regions of small (large)  perturbations of metric and spikes of non-gaussianity has an interesting  structure, which we called ``the curvaton web.'' This structure has a non-perturbative origin. 

Indeed, the non-gaussianity parameter $f_{\rm NL}(\sigma)$ takes its largest values in the regions of the universe where the  classical curvaton field $\sigma$ is small, see (\ref{v}). In the theories where the curvaton field is a real, single component field, the regions of small $\sigma$ correspond to domain walls separating large domains with $\sigma > 0$ from large domains with $\sigma < 0$  \cite{Linde:2005yw}. 

In the theory studied in the present paper, the curvaton field $\sigma$ corresponds to the radial component of a complex field $S$. In this case, the regions of small $\sigma$  form strings, reminiscent of the cosmic strings which appear due to spontaneous symmetry breaking. In our case, however, unlike in the usual cosmic string case, the curvaton strings appear in the places corresponding to the {\it minimum} of energy of the curvaton field. If one considers more complicated models, where the curvaton has $O(3)$ symmetry, instead of the domain walls and cosmic strings one will have localized objects reminiscent of global monopoles. In other words, the distribution of the peaks of non-gaussianity in the curvaton scenario has topological origin, which cannot be fully described by the standard tools of perturbation theory, such as $f_{\rm NL}$ and $g_{\rm NL}$.

\section{Discussion}

In this paper we discussed the curvaton scenario, which naturally emerges in
the simplest supergravity realization of the chaotic inflation scenario \cite%
{Kawasaki:2000yn,Kallosh:2010ug,Kallosh:2010xz}. Investigation of this
scenario consists of several parts. The main step is to find an average
value of the curvaton field $\sigma $ after a long stage of inflation. One
needs this to calculate the amplitude of perturbations of density of the
curvaton field. We performed this investigation by analyzing the growth of
the curvaton perturbations during inflation.

To conclude this investigation, one should find the ratio $r$ of the energy
of the curvaton field to the energy density of all other particles and
fields at the time of the curvaton decay. This is a complicated and
model-dependent problem, which requires study of reheating after inflation,
the decay rate of the curvaton field, and the composition of matter at the
time of the curvaton decay. In this paper, we simply treated $r$ as a free
phenomenological parameter, but one should remember that all of the issues
mentioned above should be addressed in a more detailed investigation.

We analyzed the model with the simplest quadratic inflaton potential and
with the curvaton mass given by $\alpha H^{2}+m^{2}$. Our investigation
demonstrates that if inflation is long enough, then the average value of the
curvaton contribution to the amplitude of metric perturbations, as well as
the averaged value of the non-gaussianity parameter $f_{\mathrm{NL}}$, do
not depend on initial conditions for the curvaton field. The final results
depend on the inflaton mass $m$, and on the parameter $\alpha $, which is
related to the curvature of the K{\"{a}}hler\thinspace\ manifold \cite%
{Kallosh:2010xz}. However, the locally observable parameter $f_{\mathrm{NL}}$
and the amplitude of the curvaton perturbations may take different values in
different parts of the universe and in certain cases they may significantly
deviate from their averaged values \cite{Linde:2005yw}. Moreover, the average value of the parameter $f_{\rm NL}$ can be much greater than the value $f_{\rm NL}$ in the part of the universes with an average value of the field $\sigma$. For a certain choice of parameters, the value of the non-gaussianity parameter $f_{\mathrm{NL}}$
can be in the observationally interesting range from $O(10)$ to $O(100)$.

The curvaton perturbations in our simple model have flat spectrum. This is a
consequence of degeneracy of the masses of the inflaton and curvaton field
at the end of inflation. One can change the spectral index by switching to a
theory with a different inflaton potential. This can be easily realized in
the new class of chaotic inflation models in supergravity, or by splitting
the spectrum of fluctuations of the curvaton field into two branches with
different masses \cite{Kallosh:2010ug,Kallosh:2010xz}. The last possibility
can be realized by modifying the K{\"{a}}hler\thinspace\ potential, or by
adding a term $\sim S^{3}$ to the superpotential.

Another interesting possibility is to take the inflaton mass just a little
bit smaller than $m \sim 6\times 10^{-6}$, to decrease the amplitude of the
standard inflaton perturbations. Then one may compensate for this decrease
by adding a small contribution of the curvaton fluctuations. This will
result in a smaller amplitude of tensor modes and a larger spectral index $%
n_{s}$, which would improve the agreement of the predictions of the simplest
chaotic inflation models with the WMAP data. Also, as our calculations
demonstrate, for certain values of parameters even a small contribution of
the curvaton perturbations may dramatically increase the non-gaussianity of
the combined spectrum of perturbations of metric.

Thus, whereas the curvaton models are more complicated than the single-field
inflationary models, they make the resulting scenario much more flexible,
which may be important for a proper interpretation \cite{Easson:2010uw} of
the coming observational data.

Our final comment deals with the topological features of the distribution of perturbations in the curvaton scenario. We point out that in the theory of a single-component real curvaton field, the regions of the universe with large non-gaussianity form domain walls \cite{Linde:2005yw}, reminiscent of the exponentially thick cosmic domain walls. Meanwhile in the theory of a complex curvaton field, which was studied in the present paper, the regions of large non-gaussianity form exponentially thick cosmic strings. In more complicated theories, these regions may form separate islands of large local non-gaussianity, resembling global monopoles. Since these effects have a non-perturbative, topological origin, non-gaussianity in the curvaton scenario cannot be fully described by such tools as the familiar perturbation theory parameters $f_{\rm NL}$ and $g_{\rm NL}$.

\begin{acknowledgments}
The authors are grateful to R. Bond, A. Frolov, R. Kallosh, M. Noorbala, T. Rube, M. Sasaki, and
A. Westphal for useful comments. The work by V. D. and V.M. was supported by
TRR 33 ``The Dark Universe'' and the Cluster of Excellence EXC 153 ``Origin
and Structure of the Universe.'' The work by A.L. was supported by NSF grant
PHY-0756174, by the Alexander-von-Humboldt Foundation, and by the FQXi grant
RFP2-08-19.
\end{acknowledgments}


\begin{thebibliography}{99}
 
 \bibitem{Linde:1996gt} A.~D.~Linde and V.~F.~Mukhanov, ``Nongaussian
isocurvature perturbations from inflation,'' Phys.\ Rev.\ D \textbf{56}, 535
(1997) [arXiv:astro-ph/9610219].


\bibitem{Enqvist:2001zp} K.~Enqvist and M.~S.~Sloth, ``Adiabatic CMB
perturbations in pre big bang string cosmology,'' Nucl.\ Phys.\ B \textbf{626%
}, 395 (2002) [arXiv:hep-ph/0109214].


\bibitem{Lyth:2001nq} D.~H.~Lyth and D.~Wands, ``Generating the curvature
perturbation without an inflaton,'' Phys.\ Lett.\ B \textbf{524}, 5 (2002)
[arXiv:hep-ph/0110002].


\bibitem{Moroi:2001ct} T.~Moroi and T.~Takahashi, ``Effects of cosmological
moduli fields on cosmic microwave background,'' Phys.\ Lett.\ B \textbf{522}%
, 215 (2001) [Erratum-ibid.\ B \textbf{539}, 303 (2002)]
[arXiv:hep-ph/0110096].


\bibitem{Lyth:2002my} D.~H.~Lyth, C.~Ungarelli and D.~Wands, ``The
primordial density perturbation in the curvaton scenario,'' Phys.\ Rev.\ D 
\textbf{67}, 023503 (2003) [arXiv:astro-ph/0208055].


\bibitem{Linde:1983gd} A.~D.~Linde, ``Chaotic Inflation,'' Phys.\ Lett.\ B 
\textbf{129}, 177 (1983). 

\bibitem{Linde:2005yw} A.~D.~Linde and V.~Mukhanov, ``The Curvaton Web,''
JCAP \textbf{0604}, 009 (2006) [arXiv:astro-ph/0511736].




\bibitem{Kawasaki:2000yn} M.~Kawasaki, M.~Yamaguchi and T.~Yanagida,
``Natural chaotic inflation in supergravity,'' Phys.\ Rev.\ Lett.\ \textbf{85%
}, 3572 (2000) [arXiv:hep-ph/0004243]. 



\bibitem{Davis:2008fv} S.~C.~Davis and M.~Postma, ``SUGRA chaotic inflation
and moduli stabilisation,'' JCAP \textbf{0803}, 015 (2008) [arXiv:0801.4696
[hep-ph]]. 

\bibitem{Eternal} A.~D.~Linde, ``Eternally Existing Self-reproducing Chaotic
Inflationary Universe,'' Phys.\ Lett.\ B \textbf{175}, 395 (1986).




\bibitem{Pert} V.~F.~Mukhanov and G.~V.~Chibisov, ``Quantum Fluctuation And
Nonsingular Universe,'' JETP Lett.\ \textbf{33}, 532 (1981) [Pisma Zh.\
Eksp.\ Teor.\ Fiz.\ \textbf{33}, 549 (1981)]; S.~W.~Hawking, ``The
Development Of Irregularities In A Single Bubble Inflationary Universe,''
Phys.\ Lett.\ B \textbf{115}, 295 (1982); A.~A.~Starobinsky, ``Dynamics Of
Phase Transition In The New Inflationary Universe Scenario And Generation Of
Perturbations,'' Phys.\ Lett.\ B \textbf{117}, 175 (1982); A.~H.~Guth and
S.~Y.~Pi, ``Fluctuations In The New Inflationary Universe,'' Phys.\ Rev.\
Lett.\ \textbf{49}, 1110 (1982); J.~M.~Bardeen, P.~J.~Steinhardt and
M.~S.~Turner, ``Spontaneous Creation Of Almost Scale - Free Density
Perturbations In An Inflationary Universe,'' Phys.\ Rev.\ D \textbf{28}, 679
(1983); V.~F.~Mukhanov, ``Gravitational Instability Of The Universe Filled
With A Scalar Field,'' JETP Lett.\ \textbf{41}, 493 (1985) [Pisma Zh.\
Eksp.\ Teor.\ Fiz.\ \textbf{41}, 402 (1985)]; M.~Sasaki,
``Large Scale Quantum Fluctuations in the Inflationary Universe,''
  Prog.\ Theor.\ Phys.\  {\bf 76}, 1036 (1986)." 


\bibitem{Linde:2007fr} A.~D.~Linde, ``Inflationary Cosmology,'' Lect.\ Notes
Phys.\ \textbf{738}, 1 (2008) [arXiv:0705.0164 [hep-th]].

\bibitem{book} A. D. Linde, \textit{Particle Physics and Inflationary
Cosmology} (Harwood, Chur, Switzerland, 1990); arXiv:hep-th/0503203.

\bibitem{mukhbook} V.F. Mukhanov, ``Physical Foundations of Cosmology''
(Cambridge University Press, 2005).


\bibitem{Kallosh:2010ug} R.~Kallosh and A.~Linde, ``New models of chaotic
inflation in supergravity,'' JCAP \textbf{1011}, 011 (2010) [arXiv:1008.3375
[hep-th]].


\bibitem{Kallosh:2010xz} R.~Kallosh, A.~Linde and T.~Rube, ``General
inflaton potentials in supergravity,'' arXiv:1011.5945 [hep-th]. 

\bibitem{Sasaki:2006kq}
  M.~Sasaki, J.~Valiviita and D.~Wands,
``Non-Gaussianity of the primordial perturbation in the curvaton model,''
  Phys.\ Rev.\  D {\bf 74}, 103003 (2006)
  [arXiv:astro-ph/0607627];
  K.~Enqvist and T.~Takahashi,
``Signatures of Non-Gaussianity in the Curvaton Model,''
  JCAP {\bf 0809}, 012 (2008)
  [arXiv:0807.3069 [astro-ph]].  

\bibitem{GarciaBellido:1993wn}
  J.~Garcia-Bellido, A.~D.~Linde and D.~A.~Linde,
``Fluctuations Of The Gravitational Constant In The Inflationary Brans-Dicke
Cosmology,''
  Phys.\ Rev.\  D {\bf 50}, 730 (1994)
  [arXiv:astro-ph/9312039].
 



\bibitem{Tegmark:1997in}
  M.~Tegmark and M.~J.~Rees,
``Why is the CMB fluctuation level $10^{-5}$?,''
  Astrophys.\ J.\  {\bf 499}, 526 (1998)
  [arXiv:astro-ph/9709058].


\bibitem{Garriga:2005ee}
  J.~Garriga and A.~Vilenkin,
``Anthropic prediction for Lambda and the Q catastrophe,''
  Prog.\ Theor.\ Phys.\ Suppl.\  {\bf 163}, 245 (2006)



\bibitem{Lyth:2006gd}
  D.~H.~Lyth,
``Non-gaussianity and cosmic uncertainty in curvaton-type models,''
  JCAP {\bf 0606}, 015 (2006)
  [arXiv:astro-ph/0602285].



\bibitem{Easson:2010uw} D.~A.~Easson and B.~A.~Powell, ``Optimizing future
experimental probes of inflation,'' arXiv:1011.0434 [astro-ph.CO]. 
 
\end{thebibliography}
\end{document}